\documentclass[aps,prb,nopacs,twocolumn,superscriptaddress]{revtex4-1}

\usepackage{amsfonts, amsmath, amsthm, amssymb}
\usepackage{bm}
\usepackage{graphicx}
\usepackage{color}
\usepackage[normalem]{ulem}
\usepackage[utf8]{inputenc}

\begin{document}

\title{Tuning Majorana zero modes with temperature in $\pi$-phase Josephson junctions}

\author{Umberto Borla}
\affiliation{%
 Department of Physics and Astronomy, Uppsala University, Box 516, SE-751 20 Uppsala, Sweden\\
}%
\affiliation{Department of Physics, Technical University of Munich, 85748 Garching, Germany}

\author{Dushko Kuzmanovski}
\affiliation{%
 Department of Physics and Astronomy, Uppsala University, Box 516, SE-751 20 Uppsala, Sweden\\
}%

\author{Annica M.~Black-Schaffer}
\affiliation{%
 Department of Physics and Astronomy, Uppsala University, Box 516, SE-751 20 Uppsala, Sweden\\
}%

\begin{abstract}
We study a superconductor-normal state-superconductor Josephson junction along the edge of a quantum spin Hall insulator with a superconducting $\pi$-phase across the junction. We solve self-consistently for the superconducting order parameter and find both real junctions, where the order parameter is fully real throughout the system, and junctions where the order parameter has a complex phase winding. Real junctions host two Majorana zero modes (MZMs), while phase-winding junctions have no subgap states close to zero energy. At zero temperature we find that the phase-winding solution always has the lowest free energy, which we establish being due to a strong proximity-effect into the N region. With increasing temperature this proximity-effect is dramatically decreased and we find a phase transition into a real junction with two MZMs. 
\end{abstract}

\maketitle

\section{\label{sec:intro} Introduction}
Among the many research directions in condensed matter physics, the study of topological phases of matter and their low energy degrees of freedom is currently one of the most active. Of particular interest is the search for Majorana zero modes (MZMs), elusive zero-energy quasiparticles with non-Abelian statistics that are predicted to exist as boundary states in certain topological superconductors.\cite{Qi2011,Alicea,Leijnse2012} The original, and still prototype, model for finding MZMs is the one-dimensional (1D) Kitaev model, a simple tight-binding model with spinless $p$-wave superconductivity.\cite{Kitaev2001} 

In real materials the necessary spinless $p$-wave superconductivity for MZMs is usually achieved by combining a strongly spin-orbit coupled system with a common conventional $s$-wave superconductor. Early on, it was realized that such proximity-induced conventional superconductivity generates the appropriate spinless $p$-wave symmetry in the surface states of topological insulators.\cite{FuKane2008, FuKane2009, Stanescu2010, Black-Schaffer:2011}
However, since a surface lacks natural boundaries, a junction\cite{FuKane2008, Akhmerov2009, Tanaka2009} or a superconducting vortex\cite{FuKane2008, Hosur2011} is needed in order to generate MZMs in these systems. Both superconductor-ferromagnet and superconductor-normal state-superconductor (SNS) Josephson junctions with a $\pi$-phase across the junction can host MZMs on a topological insulator surface.\cite{FuKane2008, FuKane2009} The latter case of a $\pi$-phase junction is particularly attractive since no magnetic field or proximity to a ferromagnetic material is needed, which makes for simpler experimental setups. 
Such SNS $\pi$-phase junctions can always be realized using an externally applied phase bias or alternatively, topological junctions with a $\pi$-phase ground state have also recently been proposed.\cite{Schrade15, Tkachov17, Finocchiaro17} Achieving this on both surfaces of 2D and 3D topological insulators is possible, but only in the 2D topological insulator, also called a quantum spin Hall insulator (QSHI), we find a 1D junction with the MZM being a spatially bound state and not part of a dispersive mode crossing zero energy.  

In this work we study SNS $\pi$-phase junctions on the edge of a QSHI, searching for MZMs. Intense experimental activity has already led to promising results on proximity-induced superconductivity in QSHIs, confirming signatures of both topological superconductivity and gapless Andreev bound states,\cite{Hart2014, Pribiag2015, Veldhorst2012, Bocquillon2017} such that $\pi$-phase junctions will likely soon be experimentally realized.
Despite previous results, the $\pi$-phase junction is actually deceptively simple. It can either have a superconducting order parameter that is fully real in the whole junction, resulting in a {\it real junction}, or the order parameter can wind in the complex plane between phases $0$ and $\pi$ and instead create a {\it phase-winding junction}. For topological systems this is particularly critical because in the latter case time-reversal symmetry is broken, which changes the topological class from BDI to D\cite{Schnyder2008, Tewari2012, Ardonne} and thus the topological protection of MZMs can be altered between these two types of $\pi$-phase junctions.

In a real junction belonging to the BDI class, the order parameter changes from $\Delta$ to $-\Delta$ across the junction. This results in a boundary between two different topological regions inside the junction, which always carries two MZMs.\cite{Kitaev2001, Ryu2002, FuKane2009, Tewari2012, Ardonne} These two MZMs are in fact equivalent to the single domain wall soliton present in the Su-Schrieffer-Heeger model.\cite{SuSchriefferHeeger, Takayama1980} 
On the other hand, a phase-winding junction in the Kitaev model has recently been shown to not host any MZMs.\cite{Ardonne} While this model can still topologically nontrivial, the junction itself contains {\it no} boundaries of the system, and thus {\it no} MZM can be present.

In this work we ask the very simple, yet crucial, question: How does a $\pi$-phase junction in a real QSHI behave? Is the superconducting order parameter real, as has previously just been assumed, or does it actually wind in the complex plane? And as a consequence, does a QSHI $\pi$-phase junction hosts MZMs? 
First we confirm previous results that real junctions in QSHIs contain MZMs,\cite{FuKane2009} while we find that phase-winding QSHI junctions only have non-zero subgap states and thus behaves equivalently to previous results for the Kitaev model.\cite{Ardonne} 
At zero temperature we find that the most stable $\pi$-phase junction configuration always has a phase-winding order parameter and thus no MZMs. We establish that this is due to a very strong proximity-effect into the N regions from the two S contacts. This proximity-effect makes it energetically unfavorable for the order parameter to become exactly zero in the junction, which is a requirement in a real junction.
By raising the temperature we see, however, that the proximity-effect is significantly reduced, even to the level where the real junction configuration eventually becomes energetically favorable through a first-order phase transition at a temperature $T^*$. In longer junctions the proximity-effect less strongly affects the full N region and therefore the phase transition to the real junction takes place at a lower temperature.
We further show that the phase transition from a phase-winding to a real junction leads to a sharp change in the sub-gap density of states (DOS), with no near-zero energy states in the phase-winding junction but MZMs always existing in real junctions. Moreover, we expect the $4\pi$ fractional Josephson effect to appear only at the phase transition into a real junction because it is tightly associated with the presence of MZMs.
Thus, the existence of MZMs, and their accompanying topologically protected properties, can be tuned simply by changing the temperature in QSHI $\pi$-phase junctions. 
 
The rest of the article is structured as follows. In Section \ref{sec:model} we describe the Bernevig-Hughes-Zhang (BHZ) model for a QSHI on a lattice and the superconducting state in the S regions. In section \ref{sec:results} we present our results, first focusing on the difference properties of real and phase-winding junctions, then discussing the role of proximity-effect and its temperature dependence, and finally showing how temperature can be used to tune the existence of MZMs. We end in Section \ref{sec:concl} with some concluding remarks.

\section{\label{sec:model}Model and method}
\begin{figure}[htb]
\centering
\includegraphics[width=1\linewidth]{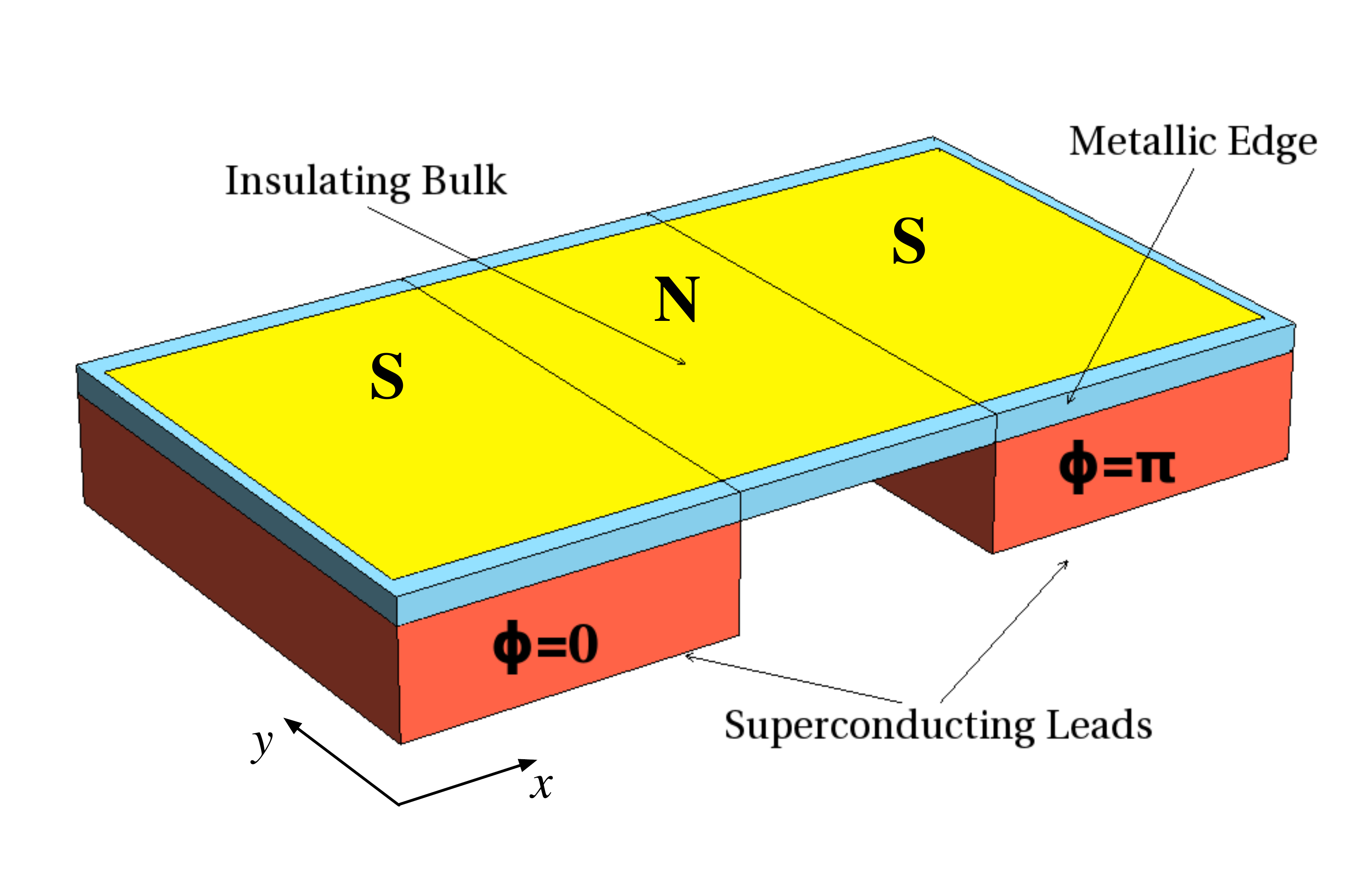}
\caption{Schematic picture of a SNS $\pi$-phase junction in a QSHI. Superconducting leads (red) are placed in contact with a QSHI (yellow bulk, blue edges) and induce superconductivity by proximity into the metallic edge states.}
\label{Draw}
\end{figure}

In this section we describe the model and method we use to study a SNS $\pi$-phase junction along the edge of a QSHI, schematically illustrated in Fig.~\ref{Draw}.

\subsection{\label{sec:modelBHZlatt} Bernevig-Hughes-Zhang lattice model}
To accurately describe the edge states of a QSHI, we employ the Bernevig-Hughes-Zhang (BHZ) model\cite{Bernevig2006} to model the full QSHI. This is an effective four-band model with two bands $|E_1\rangle$ and $|H_1\rangle$, each with a set of Kramers pair pseudospins, here represented as $\sigma = \uparrow, \downarrow$. Discretizing this model on a square lattice, the real space Hamiltonian reads\cite{Li:2016} 

\begin{align}
\label{NH}
  \mathcal{H}_0 &= \sum\limits_{i_x,i_y,i_x',i_y'} \bm{c}_{i_x,i_y}^\dagger (\hat{H}_0)_{i_x,i_y;i_x',i_y'} \bm{c}_{i_x',i_y'}  
\end{align}
with the basis spinors defined as 
\begin{equation*}
\bm{c}_{i_x,i_y} = \left(\begin{array}{c}
c_{i_x,i_y,E_1,\uparrow} \\
c_{i_x,i_y,H_1,\uparrow} \\
c_{i_x,i_y,E_1,\downarrow} \\
c_{i_x,i_y,H_1,\downarrow}
\end{array}\right),
\end{equation*}
where e.g.~$c_{i_x,i_y,E_1,\sigma}$  annihilates an electron on site $(i_x,i_y)$ in the lattice and in orbital $E_1$ and with spin $\sigma$.
The matrix
\begin{align}
  \hat{H}_0  =& \hat{T}_0\,\delta_{i_x,i_x'}\delta_{i_y,i_y'} -\hat{T}_x\,\delta_{i_x+1,i_x'}\delta_{i_y,i_y'}-\hat{T}_x^\dagger,\delta_{i_x-1,i_x'}\delta_{i_y,i_y'}\nonumber\\ 
  &- \hat{T}_y\,\delta_{i_x,i_x'}\delta_{i_y+1,i_y'} -\hat{T}_y^\dagger\,\delta_{i_x,i_x'}\delta_{i_y-1,i_y'},
\end{align}
with
\begin{align}
  \hat{T}_0 &= (-\mu+4D) + (M+4B)\,\hat{\Gamma}^5\nonumber, \\
  \hat{T}_x &= D + B\,\hat{\Gamma}^5 + (A/2i)\,\hat{\Gamma}^1,  \\
  \hat{T}_y &= D + B\,\hat{\Gamma}^5 + (A/2i)\,\hat{\Gamma}^2\nonumber,
\end{align}
where $\hat{\Gamma}^1=\sigma^x\otimes
s^z,\;\hat{\Gamma}^2=-\sigma^y\otimes 1,\;\hat{\Gamma}^3=\sigma^x\otimes
s^x,\hat{\Gamma}^4=\sigma^x\otimes s^y,\;\hat{\Gamma}^5=\sigma^z\otimes 1$, and
$\sigma^i$ and $s^i$ are the Pauli matrices acting in the $(E_1,H_1)$ and $(\uparrow,\downarrow)$ subspaces, respectively. 
Here $A$, $B$, $D$ and $M$ are parameters set by the specific material and $\mu$ is the chemical potential. The system described by the Hamiltonian Eq.~\eqref{NH} is a generally bulk insulator, but exhibits two different topological phases distinguished by the presence or absence of metallic edge states. The topologically nontrivial phase with metallic edge states occurs when $0<M/B<4$.

It is also useful for us consider a sample that is infinitely extended in the $x$ direction, i.e.~a QSHI ribbon with no SNS junction. In this case the momentum $k_x$ becomes a good quantum number. We can then Fourier transform with respect to the $x$ coordinate and obtain a semi-infinite Hamiltonian which only depends on the site index $i_y$ across the QSHI ribbon\cite{Konig:2008zz}
\begin{align}
\label{Ham1D}
  \mathcal{H}_0^{k} = 
  \sum\limits_{k_x,i_y,i_y'} \bm{c}_{k_x,i_y}^\dagger (\hat{H}_0^{k})_{i_y;i_y'}(k_x)  \bm{c}_{k_x,i_y'}  
\end{align}
with
\begin{align}
  \hat{H}_0^{k}(k_x)=\hat{M}(k_x)\delta_{i_y,i_y'}+\hat{T}\delta_{i_y,i_y'+1}+\hat{T^{\dagger}}\delta_{i_y+1,i_y'}
\end{align}
and
\begin{eqnarray}
{\hat{M}(k_x)}&=&A \sin (k_x)\hat{\Gamma}^1 -2B[2-M/2B-\cos(k_x)]\hat{\Gamma}^5\nonumber \\
&&\,\,\,\,\,\,\,\,\,\,\,\,\,\,\,\,\,\,\,\,\,\,\,\,\,\,\,\,\,\,\,\,\,-2D[2+\mu/2D-\cos
(k_x)] \nonumber\\
{\hat{T}}&=&\frac{i A}{2}\hat{\Gamma}^2+B\hat{\Gamma}^5 + D.
\end{eqnarray}

\begin{figure}
    \includegraphics[width=\linewidth]{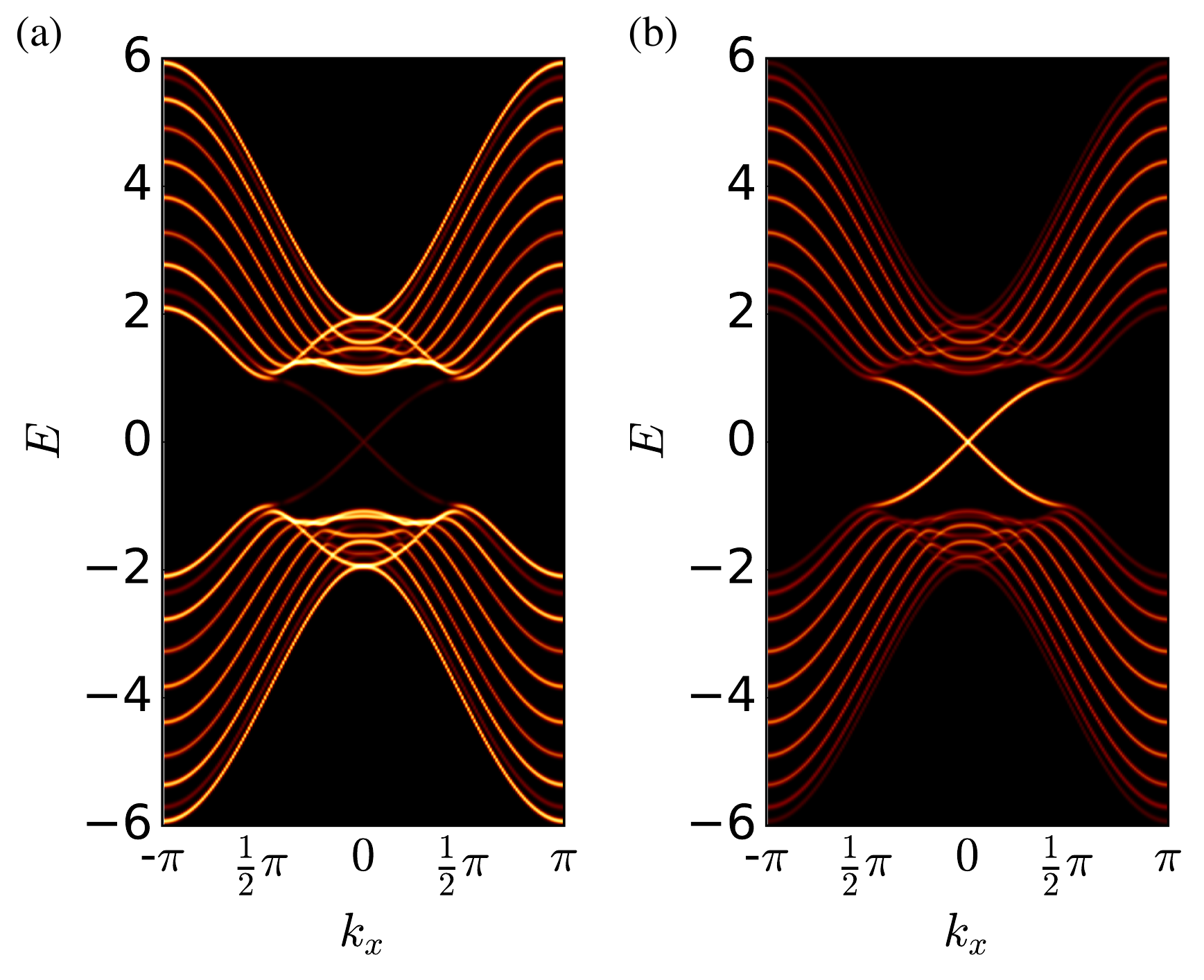} 
    \caption{Spectral function in the middle of the ribbon (a) and along the edge (b) for a semi-infinite QSHI ribbon of width $12$ in the topological phase. Parameters are given in Section~\ref{sec:modelParams}, except here $\mu =0$.}
  \label{spectral} 
\end{figure}

After solving for the eigenvalues $E_\nu$ and eigenvectors $u^\nu_{\alpha\sigma}$ of the Hamiltonian Eq.~\eqref{Ham1D}, the spectral function can be calculated as 
\begin{equation}
A(E,k_x,i_y)=\sum_{\nu,\alpha,\sigma}|u^\nu_{\alpha\sigma}(k_x,i_y)|^2\delta(E-E_\nu).
\end{equation}
One example is shown in Fig.~\ref{spectral}, where we plot the spectral function in the middle of a QSHI ribbon (a) and at the edge (b). As seen, in the topological phase the edge states are clearly visible, with their spectral weight being maximum at the edges and negligible in the bulk. Similarly, we can extract the spectral function for the full lattice Hamiltonian depending on both the $x$ and $y$-coordinates.

\subsection{\label{sec:modelSC}Proximity-induced superconductivity}
The metallic edge states make the QSHI susceptible to proximity-induced superconductivity. In the simplest model, a constant $s$-wave superconducting order parameter $\Delta$ is assumed to exist on the edge sites of the QSHI. However, when studying any heterostructure along the edge this is not sufficient, especially not in our case where the whole topological class changes critically with the behavior of the order parameter. Moreover, edge states are not truly confined to only one layer of sites, but have a finite extension also perpendicular to the edge. 
To accurately capture both of these effects, we instead assume that the effect of the external superconductors is to induce a finite attraction $U$ between the electrons in the QSHI, thus making the QSHI prone to superconductivity in the regions directly in contact with the external superconductors.
Depending on both the internal structure of the QSHI and the patterning of the external superconductor, this usually results in a spatially varying superconducting state in the QSHI, as appropriate for e.g.~ a SNS junction. For graphene SNS junctions this type of approach has been shown to give excellent experimental agreement.\cite{BlackSchaffer&Linder2010, English2016}

We here use conventional spin-singlet $s$-wave superconductors as external contacts. Their effect on the QSHI we  thus model with an effective on-site intra-orbital Hubbard attraction in the QSHI
\begin{equation}
\label{eq:HubbardInt}
\mathcal{H}_{U} = -\sum_{\bm{i} =(i_x,i_y)} U_{\bm{i}} c^{\dagger}_{\bm{i}\uparrow} c^{\dagger}_{\bm{i}\downarrow} c_{\bm{i}\downarrow} c_{\bm{i}\uparrow}.
\end{equation}
Note that the pair potential $U_{\bm{i}}\geq 0$ is site-dependent such that we capture the superconducting contact regions correctly.
Using a standard mean-field treatment this attraction is decomposed as
\begin{equation}
\label{eq:SCHam}
\mathcal{H}_{\mathrm{SC}} = \frac{1}{2}\sum_{i_{x}, i_{y}} \left[ \bm{c}^{\dagger}_{i_{x}, i_{y}} \hat{H}_{\text{SC}} (\bm{c}^{\dagger}_{i_{x}, i_{y}})^\intercal + \mathrm{H.c.} \right],   
\end{equation}
where $\check{H}_{\text{SC}}$ is diagonal in the orbital indices and spin-singlet according to
\begin{equation}
\hat{H}_\text{SC} = \text{Diag}(\Delta^{E_{1}}_{i_{x}, i_{y}},\Delta^{H_{1}}_{i_{x}, i_{y}}) \otimes (\mathrm{i} s^y).
\label{eq:Hdelta}
\end{equation}
The site- and orbital-dependent superconducting order parameter $\Delta_\mathbf{i}^\alpha$, with $\alpha=E_1$, $H_1$ labeling the orbitals, is determined self-consistently through
\begin{equation}
\label{sc}
\Delta^{\alpha}_{\mathbf{i}}=-U_{\bm{i}} \langle c^{\alpha}_{\bm{i}\downarrow}c^{\alpha}_{\bm{i}\uparrow} \rangle.
\end{equation}
Importantly, this formalism allow the superconducting order parameter to vary both with spatial position and orbital. 

We solve $\mathcal{H}= \mathcal{H}_0 +\mathcal{H}_{\mathrm{SC}}$ self-consistently within the Bogoliubov-de Gennes framework, see e.g.~Refs.~[\onlinecite{BlackSchaffer2008, BlackSchaffer&Linder2010, Kuzmanovski2016}]. In short, this entails starting with a suitable initial guess for the order parameter $\Delta^{\alpha}_{\bm{i}}$ in $\mathcal{H}$. Then by diagonalizing the Hamiltonian $\mathcal{H}$ we can compute the expectation value in Eq.~(\ref{sc}), which results in an updated value for the order parameter. 
These steps are then repeated until the difference in the order parameter between two subsequent iterations is smaller than a given small fault tolerance.
For the final converged solution, the free energy gain in the superconducting phase is computed according to
\begin{equation}
F_{S}-F_{N}=-\frac{T}{2}\sum_{\nu}\ln\left[ \frac{1 + e^{-\frac{E^\nu(\Delta)}{T}}}{1+ e^{-\frac{E^\nu(0)}{T}}}\right]+\sum_{\bm{i},\alpha}{\frac{|\Delta_{\bm{i}}^\alpha|^2}{U_{\bm{i}}}},
\end{equation}
where the first summation is over all eigenenergies $\nu$ and $T$ is the temperature in energy units (${\rm k_B} = 1$).

Since reaching self-consistency is the same as finding an order parameter configuration that creates a minimum for the free energy, the converged solution is always an energetically stable configuration. 
Note however that complications can arise when the free energy has more then one local minimum as a functional of $\Delta$. In this case, whether the iterative process converges to one or the other minimum usually depends on the initial guess. It is then necessary to calculate the free energy for all converged configurations to find the lowest energy solution.  

\subsection{\label{sec:modelParams}Choice of parameters}
To achieve a numerically tractable yet accurate model we need to carefully choose the ingoing parameters.
First, we recall the meaning of each parameter in the BHZ model. In the bulk $M$ gives the energy gap at the $\Gamma$ point and $|M-8B|$ is the highest energy eigenvalue, which means that $B$ effectively sets the bulk bandwidth. In the topological phase $A$ is the Fermi velocity of the edge states at the $\Gamma$ point, while the term containing $D$ is a band deformation that introduces an asymmetry between the electron and hole bands. 

We also need to consider how superconductivity is affected by the choice of the normal state parameters for a given pairing potential $U$. Our primary physical requirements are the following: the order parameter must be heavily suppressed in the bulk, such that superconductivity effectively only occurs along the (quasi-)1D edges. At the same time, the value of $\Delta$ on the edges needs to be large enough, such that the coherence length
$\xi=\frac{\hbar v_f}{\Delta}$ is small compared with the total size of the system. The latter condition needs to be fulfilled otherwise a superconducting state cannot develop self-consistently in the S regions.
A  well-known limitation of self-consistent numerical lattice simulations is that the dimensions of the system are heavily constrained by limitations in computational power. In this work we can use systems sizes up to 60 sites along the QSHI edge, where the junction is laid out, and $10-12$ sites wide in order to avoid edge-edge hybridization. This restricts $\xi$, which in turn generates a rather large order parameter. Still, it is important that the superconducting gap is much smaller than the bulk gap of the QSHI, which effectively contrains $M$. 

In this work we choose for easy scaling relationships $M=-2$, $B=-1$, $A=2$, and $D =0$. This clearly avoids the upper limit $M = 4B$ where the system leaves the topological phase.
For the S regions we find that $U_{\rm S}=4.8$ results in a stable superconducting state with $\Delta$ along the edge of about $0.5$, a superconducting coherence length $\xi \sim 4$ and decay length $\lambda\approx 1$ of the superconducting state into the bulk of the QSHI, which is well compatible with the normal-state parameters and system size. 
A good way to compare with previous studies, including experimental results,\cite{Konig:2008zz} is to consider the two dimensionless quantities that can be built from the ingoing normal-state parameters
\begin{equation}
m_0 = \frac{MB}{A^2}, \  r = \frac{D}{B}.
\end{equation}
With our choice of normal-state parameters we get $m_0 = 0.5$, which is only one order of magnitude larger than the physical value for a HgTe/CdTe QSHI quantum well with thickness $d=7$~nm.\cite{Konig:2008zz} 
We could further decrease $m_0$ by increasing $A$, but that results in a lower density of edge states causing superconductivity to be heavily suppressed. The only way to compensate for this would be to increase $U_{\rm S}$ to unnaturally large values. Further, by setting $D = r = 0$, we ignore the relatively unimportant asymmetry between the electron and hole bands. 

\begin{figure}[htb]
\centering
\includegraphics[width=1.03\linewidth]{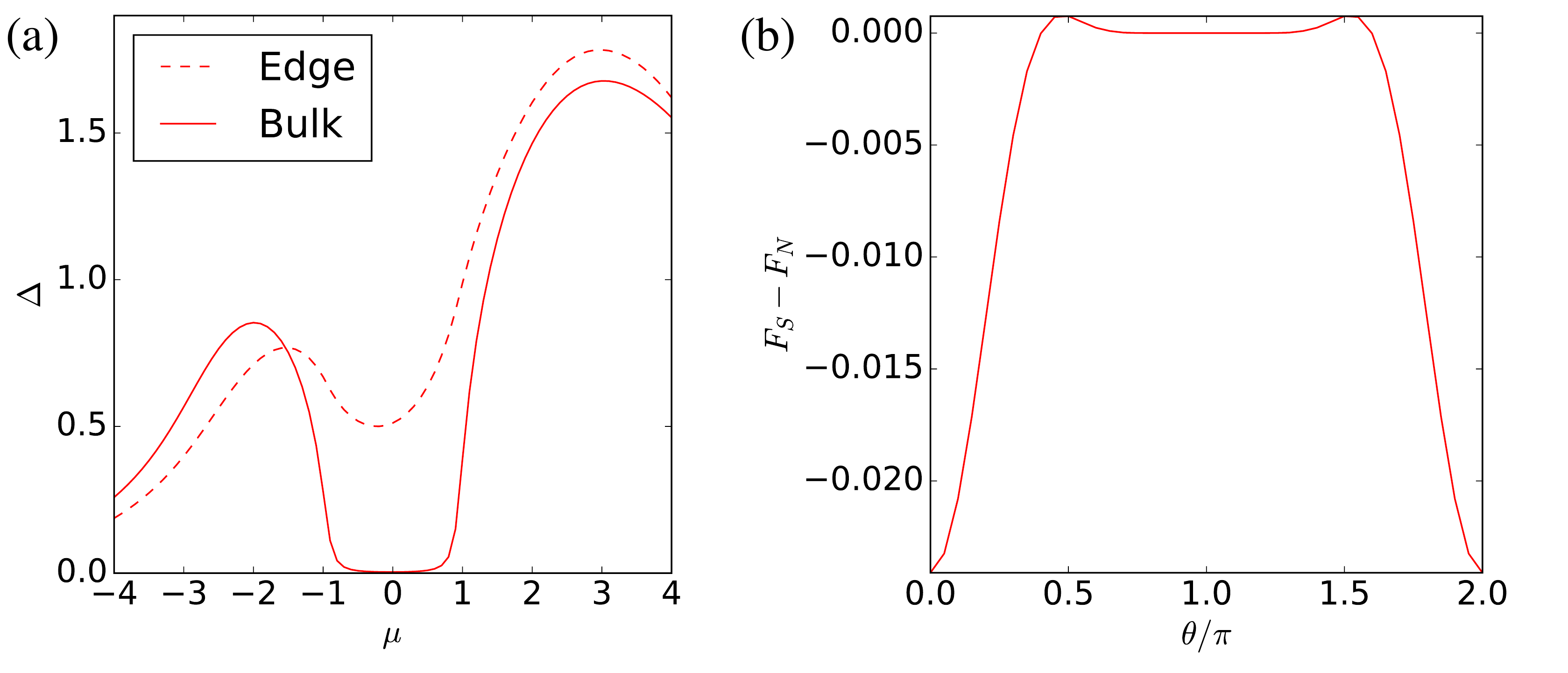}
\caption{(a) Order parameter $\Delta_{E_1}$ as a function of chemical potential $\mu$ in the bulk and along the QSHI edge. (b) Superconducting free energy $F_S-F_N$ as a function of the relative phase $\theta$ between the two order parameters $\Delta_{E_1}$ and $\Delta_{H_1}$ for a semi-infinite slab.
}
\label{Delta_F}
\end{figure}

Our choice of normal-state and $U_{\rm S}$ parameters offers a good compromise and is well within a regime to give physically relevant results. In Fig.~\ref{Delta_F}(a) we illustrate this by plotting the value of $\Delta_{E_1}$ on the edges and in the bulk as a function of the chemical potential $\mu$. As seen, a finite order parameter exists only in the bulk for $|\mu|>1$, which is where the chemical potential enters the conduction/valence bulk bands. Note here that, as long as $D=0$, we have $\Delta_{H_1}(\mu)=\Delta_{E_1}(-\mu)$ and thus for $\mu > 0$ we find $\Delta_{E_1}>\Delta_{H_1}$. To avoid unnecessary particle-hole symmetry, we usually set $\mu = 0.3$ (unless otherwise stated), which results in a stable superconducting state only along the QSHI edge. 

Since we have introduced two separate order parameters for the two orbitals $E_1$ and $H_1$, it is also necessary to establish the relative phase difference between these two. We do this by calculating the superconducting state in a semi-infinite slab where we fix their relative phase $\theta$ at each step of the self-consistent calculation. Comparing the free energy of the converged solutions in Fig.~\ref{Delta_F}(b), we find that the lowest energy configuration has a zero relative phase. Thus we are safe in assuming the same superconducting phase on $\Delta_{E_1}$ and $\Delta_{H_1}$. Notably, this is also the phase found in all our self-consistent calculations on SNS junctions, fully consistent with the result in Fig.~\ref{Delta_F}(b).

\subsection{\label{sec:junct}SNS Junction}
Having established a superconducting state along the edge, we finally create an SNS junction. For this we consider a rectangular 2D slab of the QSHI with straight edges and patterned with an SNS structure using external conventional superconductors, see Fig.~\ref{Draw}. We generally use a slab that is 12 sites wide (i.e.~perpendicular to the junction) and set $L_{\rm S} = 18$ for the length of the S regions along the junction. This length make the superconducting state in the middle of the S regions perfectly match the state found in semi-infinite slab reported in Fig.~\ref{Delta_F}(a).

We create a $\pi$-phase junction by imposing a phase difference $\pi$ between the order parameters in the two S regions, achieved by an external phase bias of the junction. We have to enforce this phase difference throughout the self-consistency procedure or it usually disappears. To do this, we fix the phase in at least the outer parts of the S regions, typically we fix the phase in a region $R = L_{\rm S}/2$. In this case the amplitude of the order parameter is allowed to relax in the whole sample, while the phase can relax self-consistently in the N region and in one half of each S region. The final converged solutions generally have a smooth profile even at the boundary between the $R$ regions and the fully self-consistent regions, and thus our procedure does not produce any unphysical effects.
Typically we choose a linear winding for the phase over the N region as a starting guess for the self-consistency process:
\begin{equation}
\Delta = \left\lbrace\begin{array}{ll} 
      \Delta_{\rm S} & , \ x \in [ 0,L_{\rm S}] \\
      \Delta_{\rm N} e^{\mathrm{i} \pi x/L_{\rm N}} &, \  x\in [ L_{\rm S},L_{\rm S}+L_{\rm N}] \\
      -\Delta_{\rm S} & ,\ x \in [ L_{\rm S}+L_{\rm N},2L_{\rm S}+L_{\rm N}].
\end{array}\right.
\end{equation}
as that generates a profile close to the converged one but still allow the system to explore both phase-winding and real junctions before reaching the converged solution.

To have a finite, albeit usually excessively small, $\Delta$ in the N region, we also need to use a finite $U_{\rm N}$ in the N region. We typically choose $U_{\rm N} = 0.1$. The superconducting transition temperature generated by this interaction alone is effectively zero because $T_c({\rm N}) \lesssim 0.001$, while $T_c({\rm S}) \approx 0.35$. This can be seen as an extremely low-temperature intrinsic superconducting state in the QSHI (not experimentally measurable), which motivates the procedure also from a physical viewpoint.
We have furthermore checked that our results are qualitatively unchanged for even smaller values of $U_{\rm N}$, including using a repulsive $U_{\rm N}$, although the latter can make convergence harder to achieve.

\section{\label{sec:results} Results}
We now turn to the results of the self-consistent calculations for SNS $\pi$-phase junctions in the QSHI. First we discuss the differences between phase-winding and real junctions, and then we address when and why junctions transition between these two cases.

\subsection{\label{sec:resultsRealvsWind}Real vs.~phase-winding junctions}
A general $\pi$-phase junction can display two different behaviors. The order parameter phase profile through the junction can interpolate more or less smoothly between $0$ and $\pi$. In such a junction the order parameter is complex and we therefore call this a {\it phase-winding} junction. Alternatively, $\Delta$ is real in the whole system. This we call a {\it real} junction and in this case the order parameter magnitude must reach zero magnitude in the middle of the junction, accompanied by a step-like change from $0$ to $\pi$ in the phase.

For the Hamiltonian $\mathcal{H}$ these two scenarios also represent different topological classes because when $\Delta$ is real $\mathcal{H}$ is invariant under time-reversal symmetry, while a complex order parameter breaks time-reversal symmetry. 
The topological classification of $\mathcal{H}$ is the same as for the 1D spinless $p$-wave superconductor state in the Kitaev chain.\cite{Kitaev2001, FuKane2009,Tewari2012, Ardonne} For a phase-winding junction the complex order parameter makes the whole system belong to symmetry class D. While the superconducting edge of a QSHI belongs to a topologically nontrivial phase, the SNS junction does not contain any boundary for this state, and therefore we should not expect any topologically protected zero modes inside the junction. The lack of zero-energy states has recently been confirmed in ideal (no change in the magnitude of $\Delta$) phase-winding junctions in the Kitaev model.\cite{Ardonne}
On the other hand, for a real junction the system belongs to class BDI. Here two different non-trivial topological phases are possible, one for $\Delta>0$ and another for $\Delta<0$. Thus a $\pi$-phase junction should automatically host a boundary between two topologically distinct phases, with the direct consequence that zero-energy boundary states should exist in the junction. These are two MZMs, which are topologically protected to appear exactly zero energy for every length $L_{\rm N}$ of the junction, as long as time-reversal symmetry is respected. 

\begin{figure}[htb]
\centering
\includegraphics[width=\linewidth]{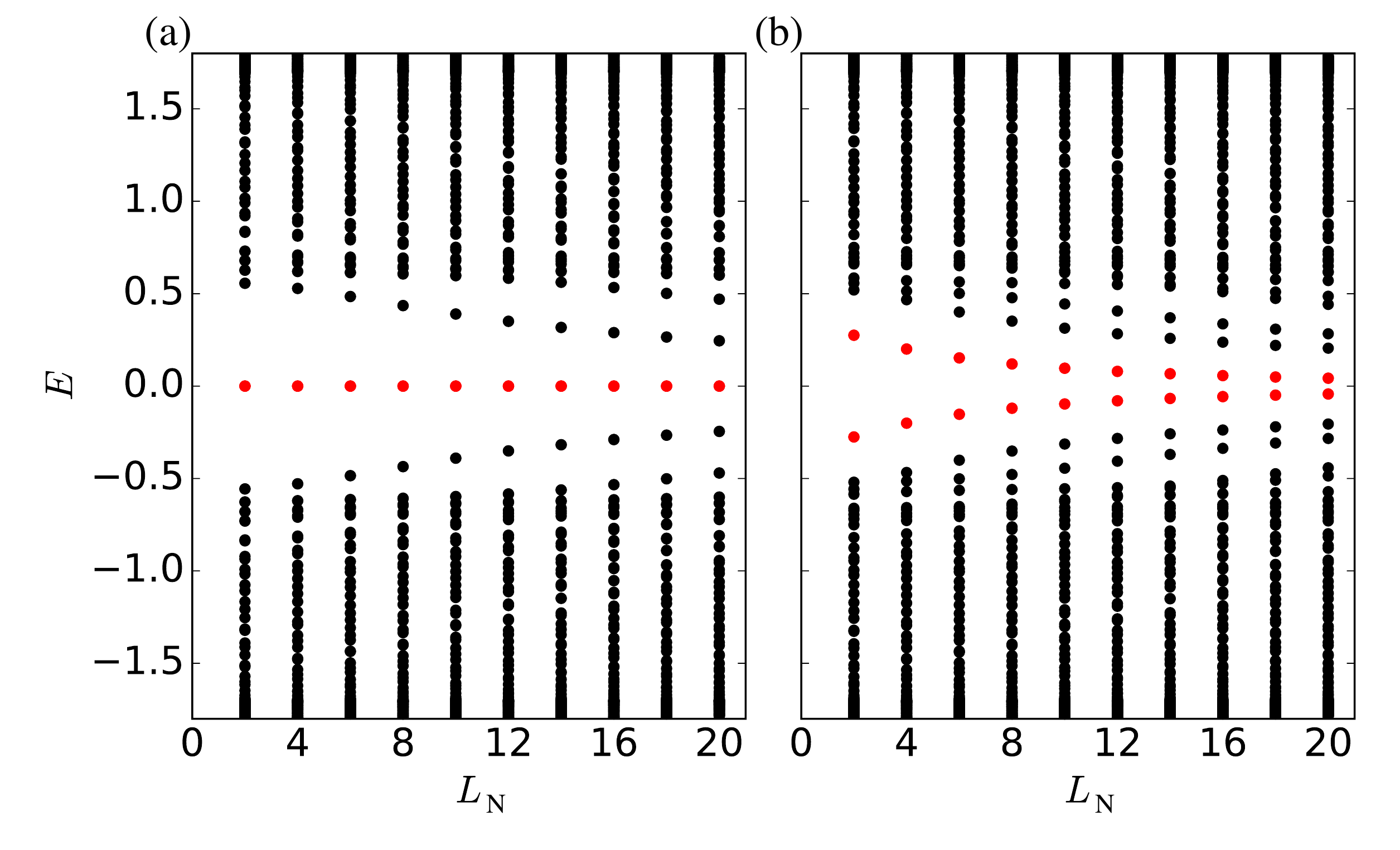}
\caption{Energy spectrum of real (a) and phase-winding (b) QSHI $\pi$-phase junctions as a function of the N region length $L_{\rm N}$. Lowest energy states are colored red for clarity.}
\label{SpectrumRealJunct}
\end{figure}

In Fig.~\ref{SpectrumRealJunct} we confirm that these predictions are correct in a realistic QSHI $\pi$-phase junction by solving fully self-consistently for the order parameter within a real (a) and phase-winding (b) junction setup of different lengths $L_{\rm N}$, from 2 to up to 20 sites.
As seen, there is always (two) zero-energy states in the real junction, independent on junction length. With increasing junction length the higher energy states move down in energy. This is expected because the proximity-induced superconductivity and its associated energy gap in the N region diminishes with increasing junction length.
To achieve this real junction we actually needed to constrain the order parameter to be equal to its absolute value, with the appropriate sign, on the whole sample at each step of the self-consistent calculation. Such a constrained calculation will not necessarily converge to the global minimum of the free energy, but at least to the local minima within the real subspace of solutions.
For the phase-winding junction we do not have to impose any other constraints on the order parameter than the phase locking in the outer $R$ part of each S region as discussed above (also needed for the real junction). Figure~\ref{SpectrumRealJunct}(b) shows the energy spectrum for this case. In short junctions we find a very large energy gap. The energy gap decreases in longer junctions as the whole energy spectrum is pushed downward in energy when the N region gets increasingly normal-like.
We have checked that these general behaviors for real and phase-winding junctions do not depend on the size $R$ of the constrained region. However, if the size of this region is increased to include up to the whole S region, we observe that the states closest to zero decreases in energy. This can be explained by the fact that a junction where the winding is constrained to a smaller region is qualitatively more similar to a real junction, which exhibits topologically protected zero-energy modes. 

\begin{figure}
\includegraphics[width=\linewidth]{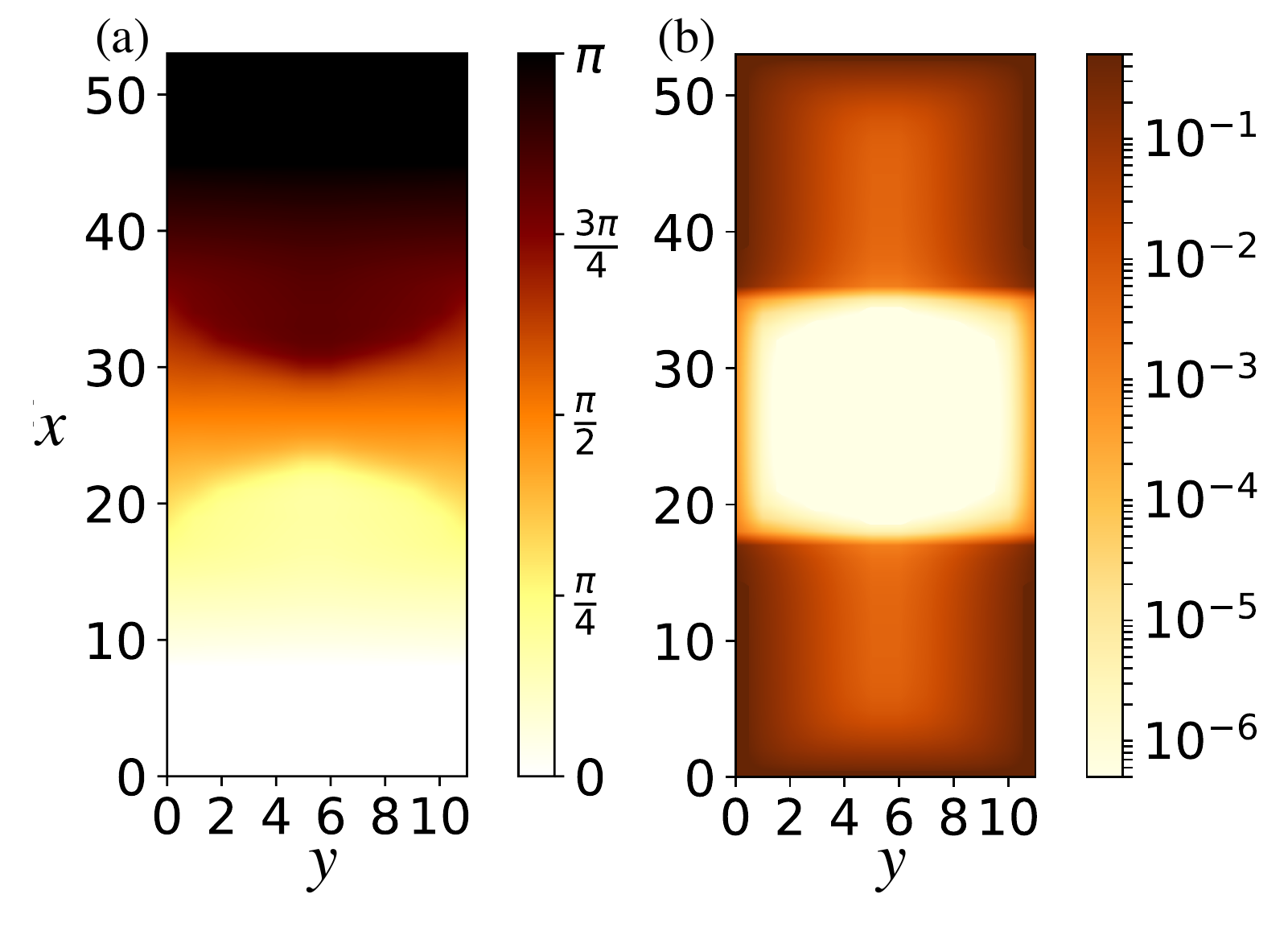} 
\caption{Phase (a) and absolute value (b) of $\Delta_{E_1}$ in the whole QSHI system for a phase-winding junction with $L_{\rm N} = 18$. Note the logarithmic scale in (b).} 
     \label{wind}
\end{figure}

Since the phase-winding junction is generated by the unconstrained calculation, this is also the true ground state of the whole system at zero temperature and $\pi$-phase bias. In Section \ref{sec:resultsFiniteT} we compare the free energies as function of temperature for both solutions to confirm this conclusion.
In Fig.~\ref{wind} we show as an example the fully self-consistent profile of the phase (a) and amplitude (b) of the superconducting order parameter over the whole QHSI for a junction of length $L_{\rm N} = 18$. Along the edges the phase winds continuously over the junction (apart from the region $R = L_{\rm S}/2$ where it is fixed), with the larger phase gradient appearing in the N region, as expected. As seen in the amplitude plot, superconductivity is heavily suppressed in the bulk of the QSHI (note the logarithmic scale).

\subsection{\label{sec:resultsProxEff}Proximity-effect in junction}
Above we found very different energy spectra for the phase-winding and real $\pi$-phase junctions. However, only the phase-winding junction was a solution of the unconstrained self-consistency equation, thus corresponding to the true ground state of the system. It is very relevant to ask whether this is always the case, or if real junctions can become the energetically favorable solution under some circumstances?
To proceed, we make the following relevant observation. The real junction necessarily needs to have the superconducting order parameter being zero at least somewhere in the N region. The phase-winding junction, on the other hand, can have a relatively large order parameter through the whole junction, as seen e.g.~in Fig.~\ref{wind}(b). Note that the value of $\Delta$ in the N region in the phase-winding case, despite being about two orders of magnitude smaller than in the S regions, is still significantly higher than can be intrinsically caused by the small attracting pair potential $U_{\rm N} = 0.1$. The magnitude of $\Delta$ in the N region is therefore almost entirely generated by a superconducting proximity-effect from the S regions. Thus, it is natural to conclude that it is primarily this proximity-effect into the N region that make the phase-winding junctions energetically preferable.

\begin{figure}[h]
\centering
\includegraphics[width=1.\linewidth]{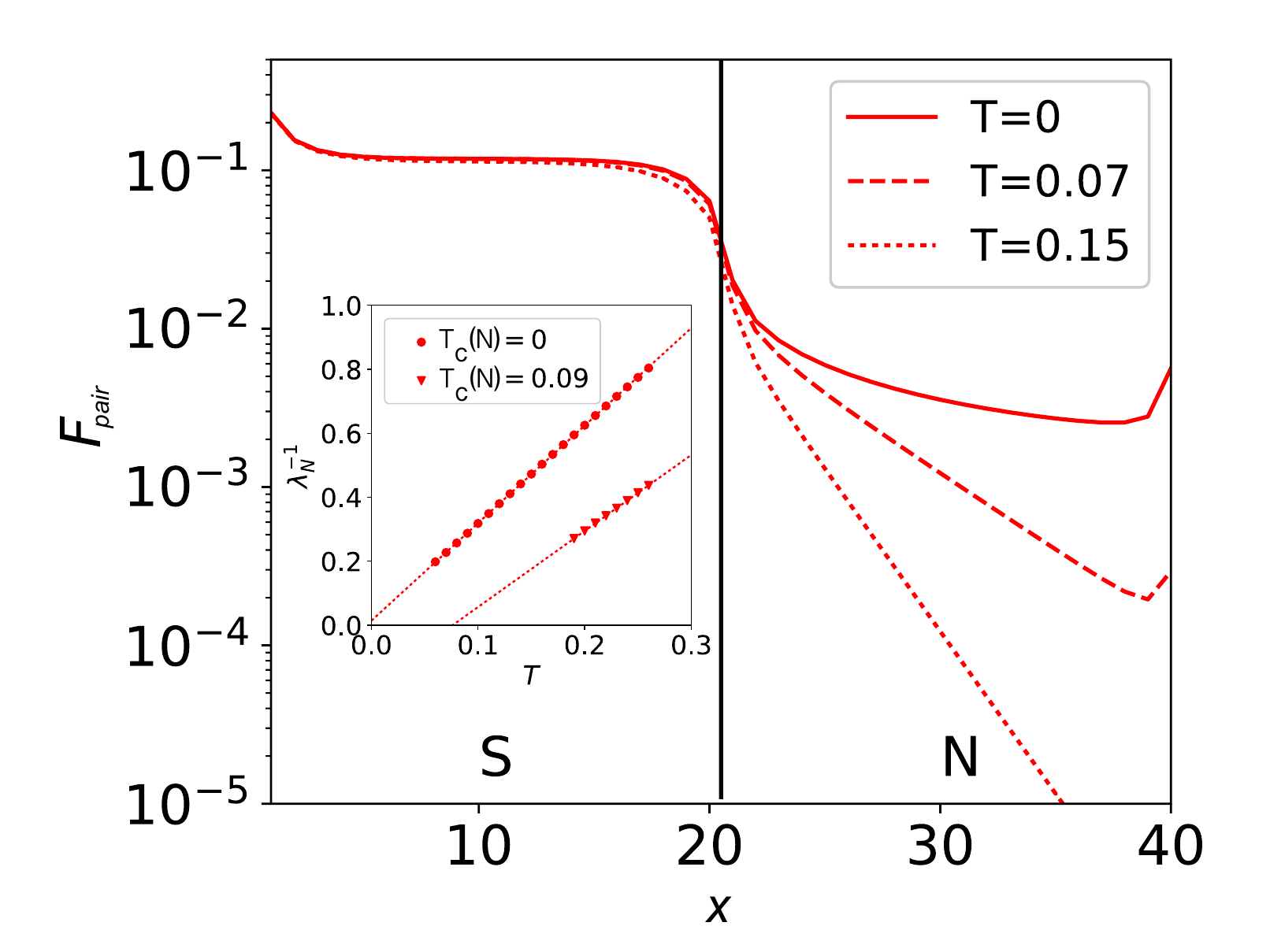}
\caption{Pair amplitude $F_{\rm pair}$ along the QSHI edge at an SN interface with $L_S=L_N=20$ at several temperatures $T$, with critical temperatures $T_c({\rm S}) \approx 0.35$ and $T_c({\rm N}) \sim 0$. Inset: Inverse decay length $\lambda_{\rm N}^{-1}$ as a function of $T$ for two different intrinsic critical temperatures in N. Here $T_c({\rm N}) = 0.09$ is achieved using $U_{\rm N} = 4$. Dashed lines are least-square fits to data points.
Data points for $\lambda_N$ can only be retrieved above a certain $T$ in the main panel, where the decay into the N region becomes exponential.}
\label{FofX}
\end{figure}

A deeper understanding of the proximity-effect from S to N along the QSHI edge can be obtained by analyzing an SN interface with the same values of $U_{\rm S}$ and $U_{\rm N}$. As a measure of the proximity-effect we use the superconducting decay length $\lambda_{\rm N}$ over which the superconducting pair amplitude $F_{\rm pair} = \langle c_{{\bm i}\downarrow} c_{{\bm i}\uparrow} \rangle$, exponentially decays from the interface and into N. The main panel in Fig.~\ref{FofX} shows the pair amplitude across the SN interface along the QSHI edge for three different temperatures, all well below $T_c({\rm S}) \approx 0.35$. For higher temperatures we see a clear exponential decay of the pair amplitude inside N, and we can then straightforwardly define an exponential decay length $\lambda_{\rm N}$.
The inset in Fig.~\ref{FofX} shows that the inverse of the decay length is a linear function of temperature. When the temperature approaches the intrinsic critical temperature of the N region there is thus a divergence of $\lambda_{\rm N}$, as also found in other systems.\cite{Covaci&Marsiglio06, Black-Schaffer:2010}

Surprisingly, we find at temperatures approaching zero that the decay of the superconducting state into N is even slower than that governed by $\lambda_{\rm N}$, despite its divergence at $T_c({\rm N})$. The decay becomes non-exponential and instead follows a power-law. There is thus always a very significant proximity-effect into the N region at low temperatures, even when the N region has an intrinsic $T_c \sim 0$. 
This much enhanced proximity-effect near $T=0$ for an SN interface is naturally also present in a SNS junction, and can there have an important impact on the junction behavior, as we show below.

\subsection{\label{sec:resultsFiniteT}Finite temperature effects}
\label{fintemp}
\begin{figure}[h]
\centering
\includegraphics[width=1.\linewidth]{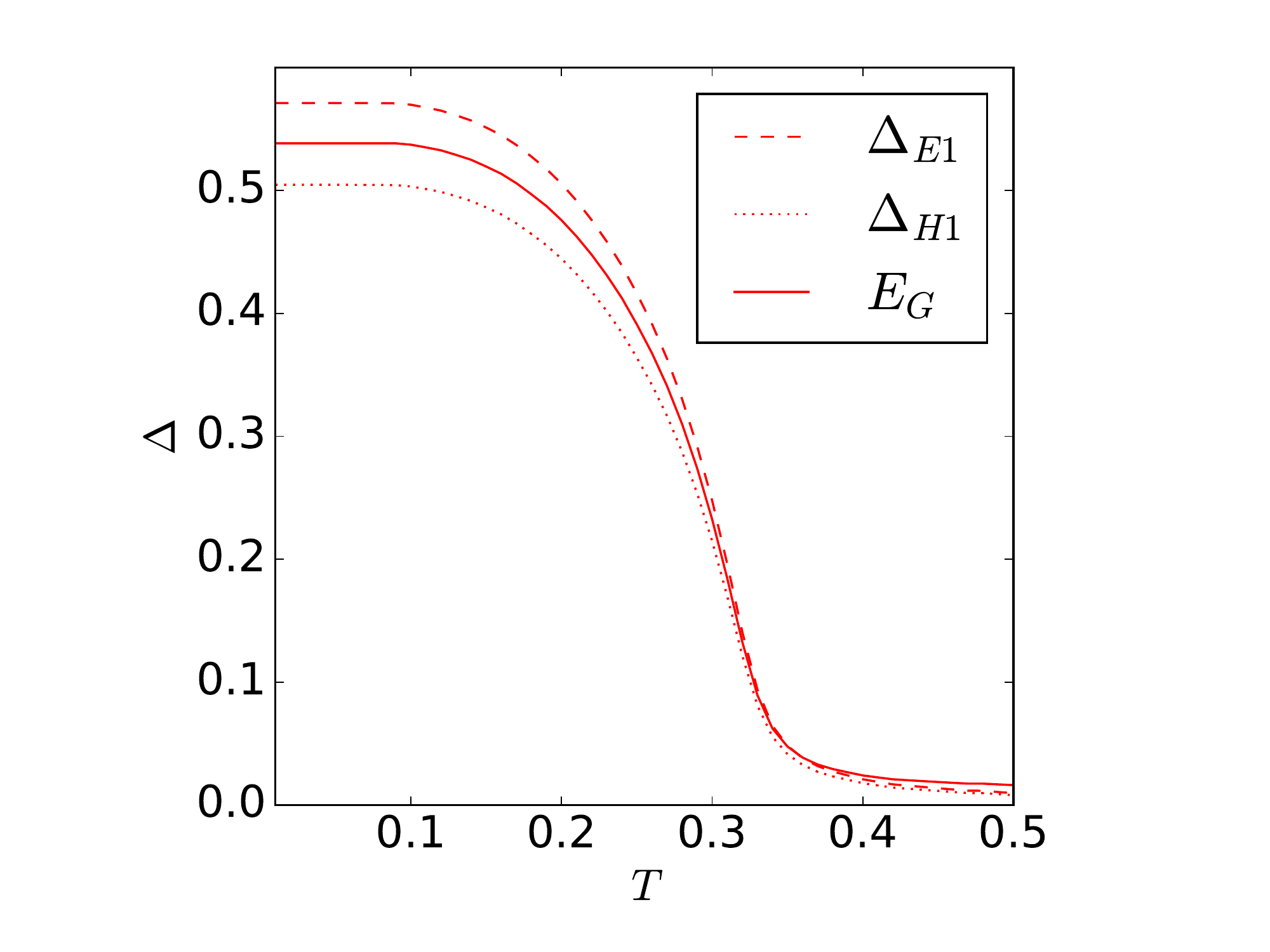}
\caption{Temperature $T$ dependence of the superconducting order parameters $\Delta_{E_1}$ and $\Delta_{H_1}$ along the QSHI edge and of the energy gap $E_G$ for a semi-infinite slab.}
\label{DofT}
\end{figure}

As we discussed in the previous section, a large proximity-effect into the N region should generally favor a phase-winding junction. We also saw that by increasing the temperature we recover a fast exponential decay of the superconducting pair amplitude inside the N region. Therefore, by raising the temperature, such that the proximity-effect in the N region is significantly reduced, it might be possible for the real junction to become the energetically most favorable solution.
As a preliminary step, we study the temperature dependence of $\Delta^\alpha$ values and the energy gap $E_G$ on the edges for a semi-infinite slab. The results are shown in Fig. \ref{DofT}. 
As seen, both order parameters and the energy gap give $T_c\approx 0.35$ and have a temperature dependence in agreement with standard BCS superconductivity.\cite{Tinkhambook} Notably, even at the highest temperatures used in Fig.~\ref{FofX}, where the pair amplitude in N is exponentially suppressed, there is no notable decay of the bulk $\Delta$ value.

To investigate the effect of temperature on a SNS $\pi$-phase junction, we need to independently find the self-consistent free energy as function of temperature for both real and phase-winding junctions. 
This is achieved by finding suitable initial guesses for which the system always falls into one these two cases. Having two such self-consistent solutions at $T = 0$, we raise the temperature in small steps, using every time the converged solution at the previous lower value of $T$ as the initial guess in the self-consistent procedure. In this way we are not only sure to find the energy minimum for each type of junction, but we also significantly speed up the convergence of the self-consistency iterations, as the initial guess is usually close to the final solution. 

\begin{figure}[htb]
    \includegraphics[width=0.9\linewidth]{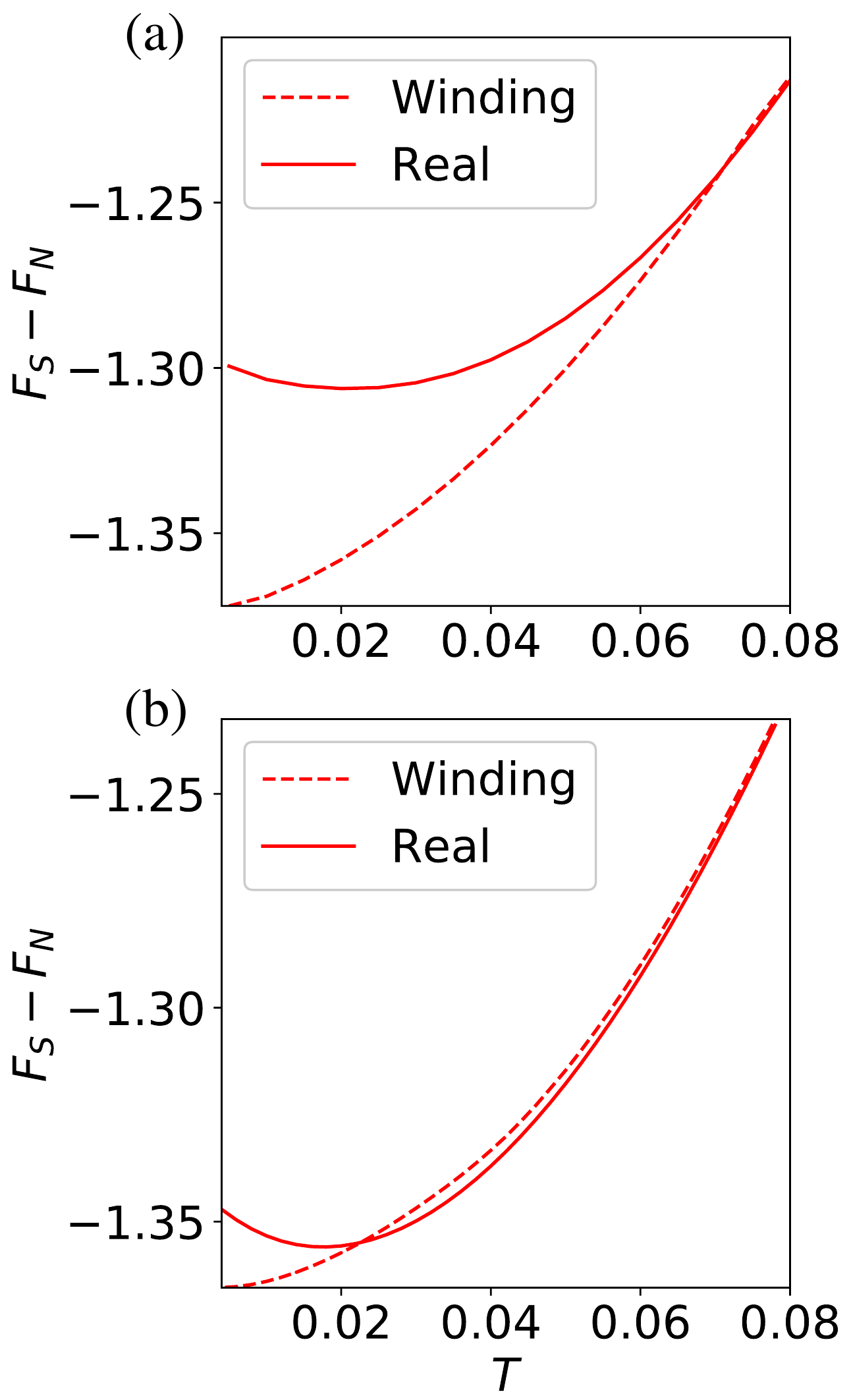} 
\caption{Superconducting free energy $F_{\rm S}-F_{\rm N}$ as a function of temperature $T$ for a short junction $L_{\rm N}=6$ (a) and long junction $L_{\rm N}=16$ (b). Note that $T_c({\rm S}) \approx 0.35$ and $T_c({\rm N}) \approx 0$.}
\label{FofT}
\end{figure}

In Fig.~\ref{FofT} we plot the superconducting free energy for the real (solid line) and phase-winding (dashed line) junctions as function of temperature for two different junction lengths $L_{\rm N}$. 
We see that at very low temperatures the phase-winding junction always has significantly lower free energy, in agreement with our previous results. However, for both junction lengths, there is a transition temperature $T^*$ where the two free energy curves intersect, resulting in the real junction becoming energetically favorable at $T>T^*$. Due to two free energy curves crossing at $T^*$ this is a first-order phase transition.
Above $T^*$, the two free energy curves do not cross again, but approach the same asymptotic value. This is due to the fact that the difference between the two cases is primarily caused by a different behavior of $\Delta$ in the middle of the $N$ region; in the real junction $\Delta$ must cross zero and thus it is generally smaller than in the winding junction case. However, at high temperatures, the proximity-effect into N is highly suppressed anyways and thus there is only an asymptotically small difference between the two free energies as $T$ increases towards the critical temperature of the whole system. 

We find that a higher temperature is required to make the order parameter in a shorter junction fully real. This can also be understood as a consequence of the proximity-effect into N. In a shorter junction the proximity-effect is naturally stronger with a relatively large $\Delta$ is present in whole N region. This superconducting pairing then requires a relatively high temperature to be suppressed and make the real junction energetically favorable. In long junctions, on the other hand, the junction midpoint is far away from the SN interfaces and the order parameter is therefore sufficiently small in the junction even at lower temperatures to favor the real junction solution. 
We further note that even for very short junctions, it is only a moderate temperature increase that is needed to achieve a real junction, especially when compared to the critical temperature of the S region, which is $T_c{\rm S} \approx 0.35$ in Fig.~\ref{FofT}. We attribute this to the exponential decay of the proximity-effect into N, with the decay length inversely proportional to temperature, which effectively suppress $\Delta$ in the N region even for moderate temperature increases.

\begin{figure}[htb]
\includegraphics[width=1.\linewidth]{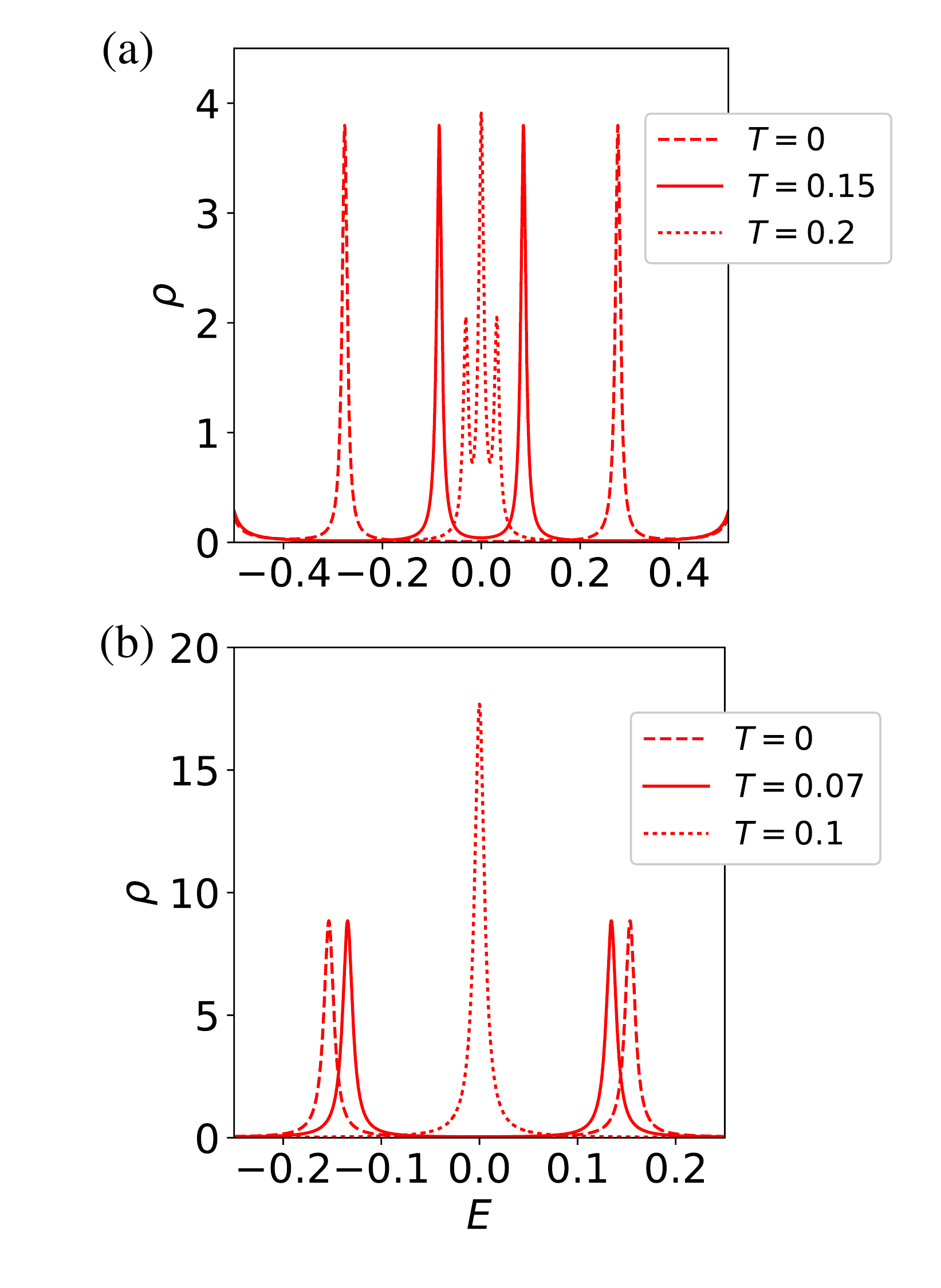} 
\caption{In-gap DOS $\rho$ for a very short junction $L_{\rm N}=2$ (a) and longer junction $L_{\rm N}=6$ (b) for several values of the temperature $T$. Dashed and solid lines are for phase-winding junctions at $T=0$ and $T \lesssim T^{*}$, respectively, while dotted lines are for real junctions at $T > T^*$.}
\label{lowEnergies}
\end{figure}

Accompanying the phase transition from a phase-winding to a real junction at $T^*$, we expect the low-energy states in the junction to abruptly change, since the real junction hosts two MZMs while the phase-winding junction has no zero-energy states, see Fig.~\ref{SpectrumRealJunct}.
In Fig.~\ref{lowEnergies} we plot the in-gap DOS for increasing temperatures for two different junction lengths $L_{\rm}$. At $T = 0$ (dashed lines) we find no low-energy states. As temperature is increased close to $T^*$, but still favoring the phase-winding junction, there is still no zero-energy states (solid lines). In fact, for longer junctions there is very little change in the energy spectrum going from $T =0$ to $T\leq T^*$, while for short junctions we find that the subgap states move towards zero energy. Eventually, at $T>T^*$ we always find two zero energy states (dotted lines).
As seen, the temperature evolution of the subgap states is more smooth in short junctions, with a gradual decrease in energy for the lowest states, while longer junctions see an abrupt transition to zero energy states at $T^*$.

\section{\label{sec:concl}Concluding remarks}
In this work we investigate the properties of a SNS $\pi$-phase junction formed along the edge of a QSHI. The superconducting state at the QSHI edge exhibit the same features as a spinless \textit{p}-wave superconducting wire, including the possibilities for different topological states. In particular, depending on the superconducting order parameter staying real or having a complex phase-winding through the $\pi$-phase junction, the topological class changes, since time-reversal symmetry is only broken in the latter case. The topological class is here not only of theoretical importance, but crucially determine if the junction contains MZMs.

Since it is impossible to {\it a priori} determine if the order parameter in a $\pi$-phase junction will be real or complex, we study a full QHSI with proximity-induced superconductivity in an SNS junction setup using a fully self-consistent approach for the superconducting order.
At $T=0$ we find the phase-winding configuration to always be energetically favorable for all junction lengths. However, as temperature increases, we discover a first-order phase transition at $T^*$ into a real junction solution. The transition temperature decreases as the junction length increases, but is even for very short junctions notably below the critical superconducting temperature for the whole system. We attribute the temperature behavior of the junction to the proximity-effect into the N region. At low temperatures there is a significant leakage of Cooper pairs into the N region from the two S contacts. This makes a real junction, where the order parameter necessarily have to be exactly zero somewhere in the junction, energetically unfavorable. However, as temperature increases, the proximity-effect is exponentially suppressed and the real junction eventually becomes the favored solution.

The phase transition between phase-winding and real junctions has a large effect on the sub-gap energy spectrum of the junction. For a real $\pi$-phase junction there are two topologically protected MZMs in the junction, while for the phase-winding junction, we find no zero-energy states. In particular, for longer junctions we find that there is a sharp transition between no low-energy states below $T^*$ and MZMs at temperatures above. Our results thus predict a very distinct temperature dependence of the DOS in the junction, directly tied to the first-order phase transition between different topological classes.
As a direct consequence there is also a strong temperature dependence on the Josephson effect in SNS junctions in QSHIs. Due to the MZM zero-energy level crossings, the real junction hosts a $4\pi$ fractional Josephson effect, while only regular $2\pi$ Josephson current behavior is present at temperatures below $T^*$.
Together, these results establish an exceptional temperature dependence of the properties in QSHI SNS junctions, due to the order parameter in the $\pi$-phase state changing the topology of the system with temperature.
\\
\acknowledgments{We thank E.~Ardonne, T.~H.~Hansson, and C.~Sp\aa nsl\"{a}tt for valuable discussions at the beginning of this project. This work was supported in part by the Carl Trygger Foundation, the Swedish Research Council (Vetenskapsrådet) (Grant No. 621-2014-3721), and the Knut and Alice Wallenberg Foundation
through the Wallenberg Academy Fellows program.}

\end{document}